\documentclass[runningheads]{llncs}
\usepackage{graphicx}
\usepackage{comment}
\usepackage{amsmath,amssymb} 
\usepackage{color}
\usepackage{amssymb}
\usepackage{booktabs}
\usepackage{multirow}
\usepackage{siunitx}
\usepackage{subcaption}
\usepackage{url}
\usepackage[width=122mm,left=12mm,paperwidth=146mm,height=193mm,top=12mm,paperheight=217mm]{geometry}

\begin{document}
\pagestyle{headings}
\mainmatter
\def\ECCVSubNumber{4749}  

\newcommand{\revwei}[1]{\textcolor{black}{#1}}
\newcommand{\revnao}[1]{\textcolor{black}{#1}}
\newcommand{\revdan}[1]{\textcolor{black}{#1}}
\newcommand{\rev}[1]{\textcolor{black}{#1}}

\title{Guided Deep Decoder: Unsupervised Image Pair Fusion} 

\titlerunning{Guided deep decoder}
%
\author{Tatsumi Uezato\inst{1}\orcidID{0000-0002-8264-201\textrm{X}} \and
Danfeng Hong\inst{2,3}\orcidID{0000-0002-3212-9584} \and
Naoto Yokoya\inst{4,1}\orcidID{0000-0002-7321-4590} \and
Wei He\inst{1}\orcidID{2222--3333-4444-5555}}
\authorrunning{T. Uezato et al.}
%
\institute{RIKEN AIP, Tokyo, Japan \and
German Aerospace Center, Wessling, Germany \and
Univ. Grenoble Alpes, CNRS, Grenoble INP, GIPSA-lab, France \and
The University of Tokyo, Tokyo, Japan\\
\email{\{tatsumi.uezato,naoto.yokoya,wei.he\}@riken.jp}, \email{\{danfeng.hong\}@dlr.de}}
\maketitle

\begin{abstract}
The fusion of input and guidance images that have a tradeoff in their information (e.g., hyperspectral and RGB image fusion or pansharpening) can be interpreted as one general problem. However, previous studies applied a task-specific handcrafted prior and did not address the problems with a unified approach. To address this limitation, in this study, we propose a guided deep decoder network as a general prior. The proposed network is composed of an encoder-decoder network that exploits multi-scale features of a guidance image and a deep decoder network that generates an output image. The two networks are connected by feature refinement units to embed the multi-scale features of the guidance image into the deep decoder network. The proposed network allows the network parameters to be optimized in an unsupervised way without training data. Our results show that the proposed network can achieve state-of-the-art performance in various image fusion problems.
\end{abstract}
\begin{figure*}[h]
        \centering
        \includegraphics[width=\linewidth]{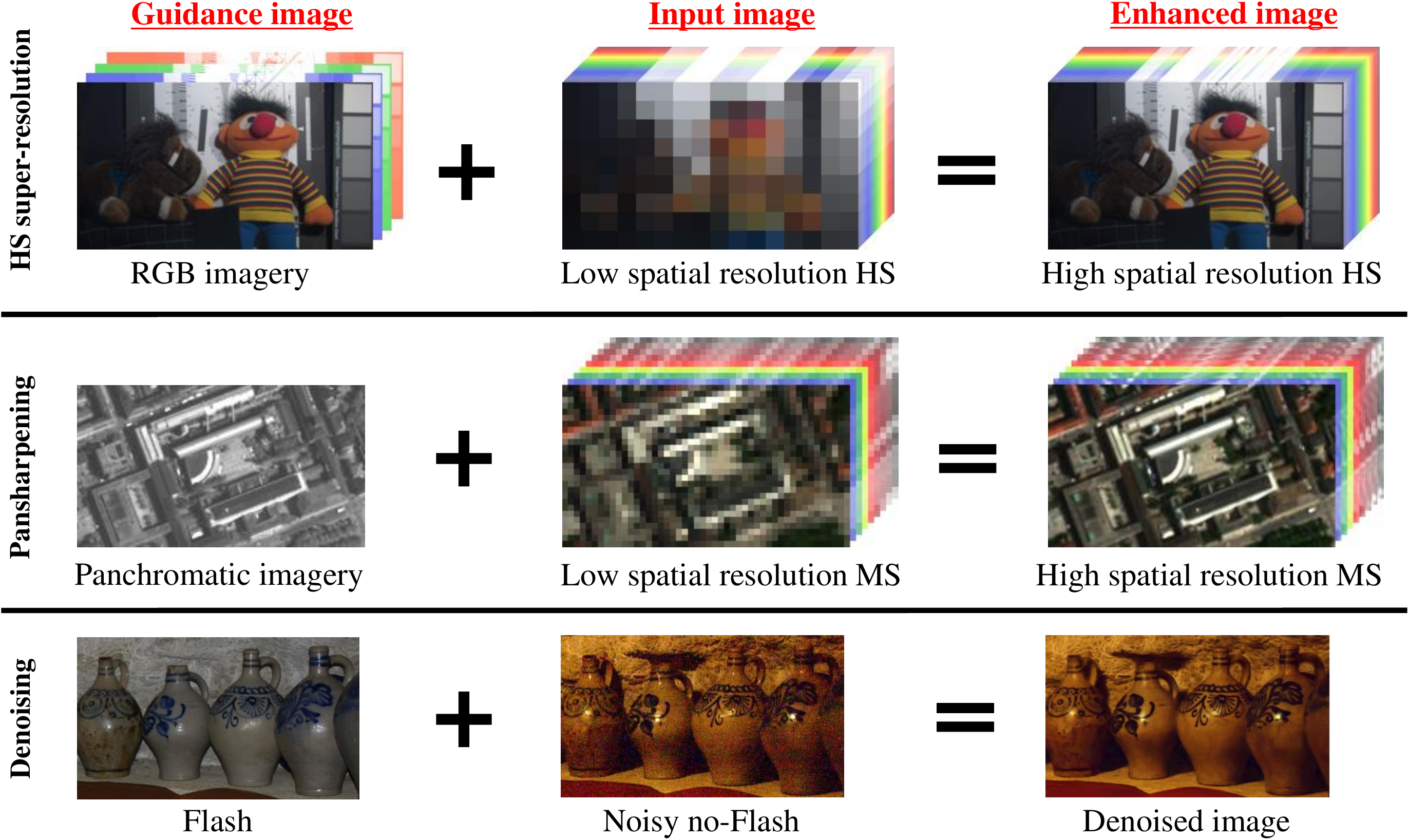}
        \caption{Illustration of image pair fusion of the same modality.}\label{fig:intro}
\end{figure*}
\section{Introduction}
Some image fusion tasks address the fusion of image pairs in the same modality. The tasks consider a pair of images that capture the same region but have a tradeoff between the two images (Fig.~\ref{fig:intro}). For example, a low spatial resolution hyperspectral (LR-HS) image has greater spectral resolution at lower spatial resolution~\cite{yokoya2017hyperspectral}. However, an RGB image acquires much lower spectral resolution at higher spatial resolution. Likewise, panchromatic and multispectral (MS) images have a tradeoff between spatial and spectral resolution~\cite{vivone2014critical}. No-flash images capture ambient illumination, but are very noisy, while flash images capture artificial light, but are less noisy~\cite{Petschnigg2004}. Image fusion enables an image that overcomes the tradeoff to be generated. Hyperspectral super-resolution or pansharpening aims to generate a high resolution (HR) HS or MS image. The denoising of a no-flash image with a flash image can be also interpreted as a special case of image fusion.

Although these tasks share a common goal (i.e., enhancing input images with the help of guidance images), the tasks have been studied independently. This occurs because a different handcrafted prior is considered to incorporate the specific property of an output image. In HS super-resolution, a prior exploiting the low-rankness of HS has been extensively used~\cite{yokoya2012coupled,Dian2018b,Lanaras2015}. In pansharpening, a prior representing a spatial smoothness has been considered~\cite{Palsson2014}. The denoising task assumes that the spatial structure of a restored image is similar to that of a guidance image~\cite{Petschnigg2004}. \rev{While these handcrafted priors share the same goal, the priors need to be designed for each task to exploit the specific properties of data.} It is highly desirable to develop a prior applicable to various image fusion problems.

Deep learning (DL) approaches avoid the assumption of explicit priors for each specific task. Although network architectures themselves need to be handcrafted, properly designed network architectures have shown to solve various problems~\cite{ronne2015unet,He2016resnet}. Most DL approaches rely on training data. However, for pan-sharpening and hyperspectral super-resolution, it is difficult to collect a large size of training data including reference (i.e., HR-HS or HR-MS) because of the cost or hardware limitation. Thus, previous studies~\cite{xie2019multispectral,Scarpa2018} have frequently used synthetic data for training, which may have limited generalization performance. In addition, different sensors provide different spectral response functions. Networks trained on data acquired by a particular sensor may not work well on new data acquired by a different sensor.

A natural question arises: is it possible to use DL approaches without training data? Ulyanov \textit{et al.}~\cite{Ulyan2018} have shown that network architectures have inductive bias and can be used as deep image prior (DIP) without any training data. This intriguing property of DIP has been successfully used for various problems~\cite{Gandelsman_2019_CVPR,Yokota_2019_ICCV,Sidorov2019}. In~\cite{Ulyan2018}, the guided denoising task of flash and no-flash image pair has been addressed using a no-flash image as an input and a flash image as an output. Although this approach can be potentially used to address the problems shown in Fig.~\ref{fig:intro}, the network architecture does not fully exploit the semantic features or image details of a guidance image. It is still unclear how the network architecture is conditioned on the features of a guidance image. Although DIP has great potential, the uncertainties limit DIP to achieve state-of-the-art (SOTA) performance in various image fusion problems. 

As discussed above, previous studies face two major problems (task-specific handcrafted priors and requirement of training data) to address various image fusion problems in a unified framework. In this study, we propose a new network architecture, called a guided deep decoder (GDD), that overcomes the problems and can achieve SOTA performance in different image fusion problems. Specifically, the proposed network architecture is composed of two networks where one encoder-decoder network is designed to extract multi-scale semantic features from a guidance image, while another deep decoder network generates an output image from random noise. The two networks are connected by feature refinement units incorporating attention gates to embed the multi-scale features of the guidance image into the deep decoder network.

The contributions of this paper are as follows. (1) We propose a new unsupervised DL method that does not require training data and can be adapted to different image fusion tasks in a unified framework. We achieve SOTA results for various image fusion problems. (2) We propose a new network architecture as a regularizer for unsupervised image fusion problems. The attention gates used in the proposed architecture guide the generation of an output image using the multi-scale semantic features from a guidance image. The guidance of the multi-scale features can lead to an effective regularizer for an ill-posed optimization problem.
\section{Related work}
Most of the previous works have independently addressed one of the image fusion problems shown in Fig.~\ref{fig:intro}, although the common goal is to generate an image that overcomes the tradeoff. This study focuses on the data acquired in the same modality and is different from the image fusion problems of different modalities where the sensor captures different physical quantities (e.g., fusion of RGB images and depth maps~\cite{Lutio_2019_ICCV}). To address the ill-posed fusion problems, similar approaches have been developed for different image fusion tasks.

\textbf{Classical approach}: The classical approach is to specifically design a handcrafted prior for each task. For example, handcrafted priors exploiting the low-rankness or sparsity of HS have been developed for HS and MS image fusion problems~\cite{yokoya2012coupled,kawakami2011,Kwon2015,Qi_BFUSE}. In panchromatic and MS image fusion, the handcrafted priors, which assume that the spatial details of PAN are similar to those of MS, have been widely used~\cite{Liu2018pan,Palsson2014,Fu2019pan,Chen2014}. In addition, flash and no-flash image fusion uses a prior that promotes similar spatial details between the paired image~\cite{Petschnigg2004}. The classical approach can reconstruct an enhanced image without any training data by explicitly assuming prior knowledge. However, the priors designed for a specific task may not be effective when they are applied to other tasks. In addition, an optimization method needs to be tailored for a different prior.

\textbf{Supervised DL approach}: DL methods that use training data have recently achieved SOTA performance in different image fusion problems. DL methods are usually built upon a popular network (e.g.,~\cite{ronne2015unet,He2016resnet}). In the HS and RGB image fusion, DL methods use LR-HS and RGB images as an input and an HR-HS image as an output and learn the mapping function between the inputs and the output~\cite{xie2019multispectral,Dian2018}. Similarly, in pansharpening, the methods consider panchromatic and LR-MS images as an input and HR-MS as an output and learn the mapping function~\cite{Scarpa2018,Wei2017,Yang2017}. As long as training data are available, DL methods can be potentially applied to different image fusion problems in a unified framework by slightly changing the network architecture or the loss function. However, it may be difficult to acquire training data, including reference data, for HS or MS images because of the cost or hardware limitation.

\textbf{Unsupervised DL approach}: To bridge the gap between the classical and supervised DL approaches, an unsupervised DL approach has been considered in some studies. The unsupervised DL methods have been developed to address the HS and RGB image fusion problem~\cite{qu2018unsupervised,Fu2019hyper}. In~\cite{qu2018unsupervised,Fu2019hyper}, the network architecture has been specifically designed to exploit the property of the HS image and different handcrafted priors have been combined to achieve optimal performance. However, it may not achieve SOTA performance in other tasks because of the specifically designed network and handcrafted priors. DIP that can apply DL in an unsupervised way has been recently developed by~\cite{Ulyan2018} and has been applied for a variety of problems~\cite{Gandelsman_2019_CVPR,Yokota_2019_ICCV,Sidorov2019}. Although DIP can be potentially applied for various image fusion problems, it has not been explored yet. The simple application of DIP cannot achieve SOTA performance in different image fusion tasks, which is shown in the following experiments. Our study borrows the idea of DIP and proposes a robust network architecture that achieves SOTA performance in these tasks. 
\section{Methodology}
\subsection{Problem formulation}
Let us denote a low resolution or noisy input image $\mathbf{Y} \in \mathbb{R}^{C \times w \times h}$ and a guidance image $\mathbf{G} \in \mathbb{R}^{c \times W \times H}$ where $C$, $W$, and $H$ represent the number of channels, the image width, and the image height, respectively. When considering HS super-resolution or pansharpening, $w \ll W$, $h \ll H$, and $c \ll C$. In the unsupervised image fusion problem, the corresponding output $\mathbf{X} \in \mathbb{R}^{C \times W \times H}$ can be estimated by solving the following optimization problem:
\begin{equation}
	\min\limits_{\mathbf{X}}\mathcal{L}\left(\mathbf{X}, \mathbf{Y}, \mathbf{G}\right) + \mathcal{R}\left(\mathbf{X}\right),
\end{equation}
where $\mathcal{L}$ is a loss function that is different for each task. Because the problem is ill-posed, existing methods commonly add the handcrafted regularization term $\mathcal{R}$. However, the task-specific regularization term (e.g., low-rank property of HS images) cannot be easily applied to other tasks. Instead of using the handcrafted regularization terms, DIP estimates $\mathbf{X}$ using a convolutional neural network (CNN)-based mapping function as:
\begin{equation}
	\mathbf{X} = f_{\theta}\left( \mathbf{Z}\right),
\end{equation}
where $f_{\theta}$ represents the mapping function with the network parameters $\theta$, $\mathbf{Z}$ is the input representing the random code tensor. The optimization problem can be rewritten as:
\begin{equation}
	\min\limits_{\theta}\mathcal{L}\left(f_{\theta}\left(\mathbf{Z}\right), \mathbf{Y}, \mathbf{G}\right).
\end{equation}
In this formulation, only one input image $\mathbf{Y}$ and a guidance image $\mathbf{G}$ are used for the optimization problem; thus, training data are \textit{not} required. $\mathbf{X}$ is regularized by the implicit prior of the network architecture. Different types of architectures can lead to different regularizers. The architecture that effectively incorporates multi-scale spatial details and semantic features of the guidance image can be a powerful regularizer for the optimization problem. In the following section, we propose a new architecture, called the guided deep decoder, as a regularizer that can be used for various image fusion problems.
\begin{figure*}[!t]
        \centering
        \includegraphics[width=\linewidth]{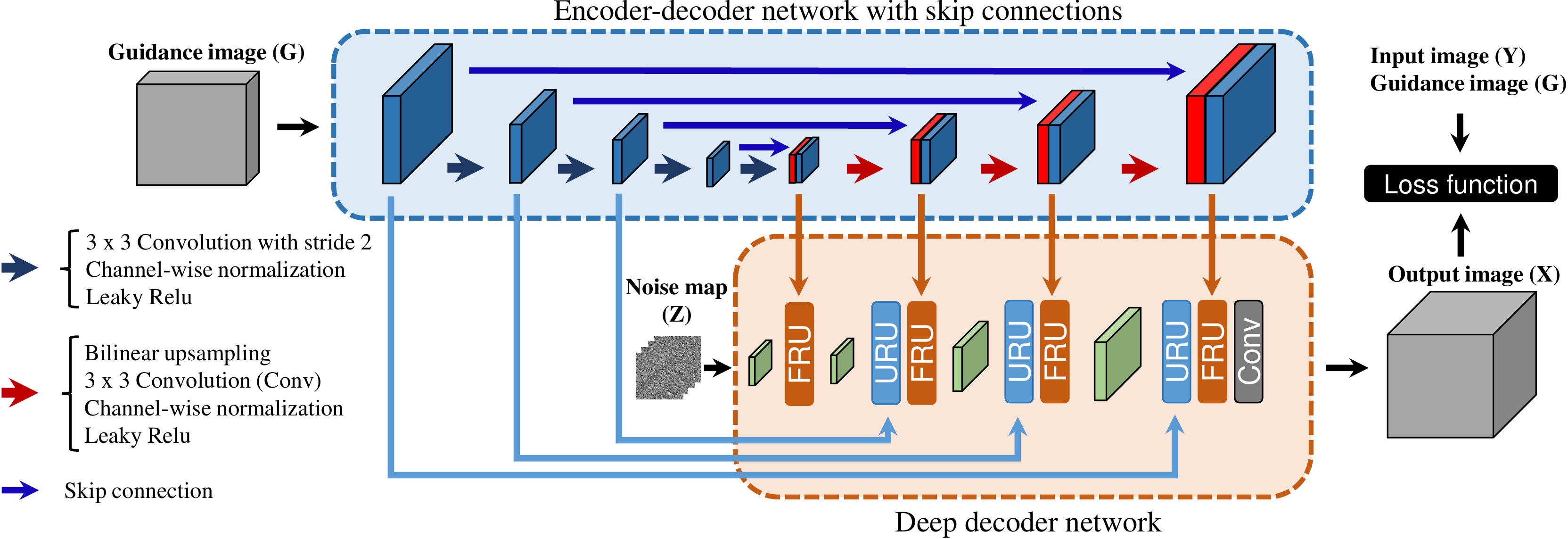}
        \caption{The structure of a guided deep decoder. The semantic features are extracted from the guidance image by the U-net like encoder-decoder network. The blue layers represent the features of the encoder. The red layers represent the features of the decoder. The green layers represent the features of the deep decoder network. The semantic features of G are used to guide the features of the deep decoder in the upsampling and feature refinement units (URU and FRU).}\label{fig:method}
\end{figure*}
\begin{figure*}[!t]
        \centering
        \includegraphics[width=.7\linewidth]{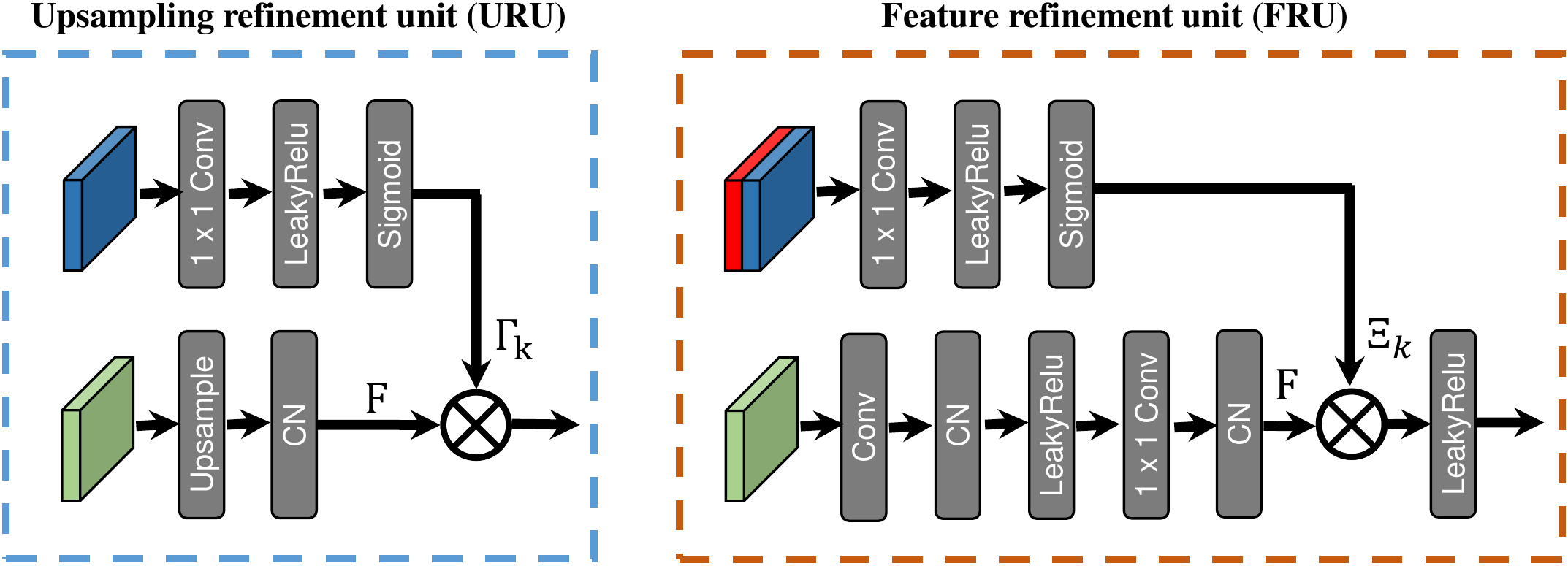}
        \caption{The structure of upsampling and feature refinement units.}\label{fig:unit}
\end{figure*}
\subsection{Guided deep decoder (GDD)}
GDD is composed of an encoder-decoder network with skip connections and a deep decoder network, as shown in Fig. \ref{fig:method}. The encoder-decoder network is similar to the architecture of U-net~\cite{ronne2015unet} and produces the features of a guidance image at multiple scales. The multi-scale features represent hierarchical semantic features of the guidance image from low to high levels. The semantic features are used to guide the parameter estimation in the deep decoder. Let $\mathbf{\Gamma}_k$ denote the features of the encoder at the $k$th scale, $\mathbf{\Xi}_k$ denotes the $k$th-scale features in the decoder part of the encoder-decoder network. The mapping function is conditioned on the multi-scale features as $f_{\theta}\left( \mathbf{Z}\right|\mathbf{\Gamma}_1,\cdots,\mathbf{\Gamma}_K,\mathbf{\Xi}_1,\cdots,\mathbf{\Xi}_K)$. The multi-scale features are incorporated in the deep decoder by the two proposed units shown in Fig.~\ref{fig:unit}.
\subsubsection{Upsampling refinement unit (URU).}
Upsampling is a vital part of DIP~\cite{Chakrabarty2019}. Bilinear or nearest neighbor upsampling promotes piecewise constant patches or smoothness across all channels~\cite{heckel2019}. However, the prior is too strong to recover exact spatial structures or boundaries of an image. Although this problem is alleviated using skip connections, the spatial details of a guidance image are still lost in the features of the decoder. URU incorporates an attention gate for weighting the features derived after upsampling and channel-wise normalization (CN) in the deep decoder. The features from the guidance image are gated by a $1 \times 1$ convolution (Conv), a leaky rectified linear unit (LeakyRelu), and a sigmoid activation layer (Sigmoid) to preserve the spatial locality of the features and generate the conditional weights. Given the features of the deep decoder $\mathbf{F}$, the transformation is carried out as:
\begin{equation}
	\textbf{URU}\left( \mathbf{F}|\mathbf{\Gamma}_k \right) = \mathbf{F} \otimes \mathbf{\Gamma}_k,
\end{equation}
where $\otimes$ represents the element-wise multiplication. Note that the dimensions of $\mathbf{F}$ and $\mathbf{\Gamma}_k$ are the same at each scale. Both channel-wise and spatial-wise conditional weights are considered in URU.
\subsubsection{Feature refinement unit (FRU).}
FRU is different from URU in that the features of the deep decoder are weighted by the high-level semantic features of the guidance image. FRU promotes the semantic alignment with the features of the guidance image, while URU promotes similar spatial locality. Using an attention gate, the high-level features are gated by a $1 \times 1$ convolution, a leaky rectified linear unit, and a sigmoid activation layer to generate the conditional weights. FRU transforms the features of the deep decoder as follows:
\begin{equation}
	\textbf{FRU}\left( \mathbf{F}|\mathbf{\Xi}_k \right) = \mathbf{F} \otimes \mathbf{\Xi}_k.
\end{equation}
Note that the dimensions of $\mathbf{F}$ and $\mathbf{\Xi}_k$ are the same at each scale. The features of the deep decoder are weighted in URU and FRU, which leads to a deep prior that can more explicitly exploit the spatial details or semantic features of the guidance image than DIP.
\subsection{GDD and existing network architectures}
GDD is closely related to existing network architectures. URU and FRU in GDD generate multiplicative transformation parameters from the guidance image for the spatial and channel-wise feature modulation. A similar feature modulation has been also used in~\cite{wang2018sftgan,fu2018dual,Islam_2017_CVPR,Park_2019_CVPR,Liu2019}. In~\cite{wang2018sftgan}, affine transformation parameters have been generated from segmentation probability maps for the feature modulation to achieve more realistic textures in image super-resolution. The affine transformation has been also considered in the style transfer~\cite{Huang2017}. Although the affine transformation considers the scaling and bias values, URU and FRU consider only the scaling values because we find that similar results can be obtained at lower computational cost for unsupervised optimization problems.

The conditional weights can be also interpreted as attention layers across all channels. In~\cite{fu2018dual,Oktay2018,Islam_2017_CVPR}, attention gates are incorporated to refine the spatial details and highlight salient features. The conditional attention weights are generated from a label map for semantic image synthesis~\cite{Liu2019}. GDD is closely related to the conditional attention weights in that it uses the multi-scale features from a guidance image to generate the conditional attention weights. However, all of the aforementioned studies consider and require a large size of training data. The network architectures have not been fully explored as a regularizer for the unsupervised optimization problems. Our study is different from previous studies in that it uses the network architecture as a regularizer to solve a variety of unsupervised image fusion problems.
\subsection{Loss Function}
The loss function is different for each task. In this section, the loss functions used for HS super-resolution, pansharpening, and denoising are discussed.
\subsubsection{HS super-resolution.}
When fusing RGB and HS images, the loss function is usually designed to preserve the spectral information from the HS image while keeping the spatial information from the RGB image. For simplicity, the matrix forms of $\mathbf{X}, \mathbf{Y}, \mathbf{G}$ are denoted as $\mathbf{\tilde{X}} \in \mathbb{R}^{C \times WH}$, $\mathbf{\tilde{Y}} \in \mathbb{R}^{C \times wh}$, and $\mathbf{\tilde{G}} \in \mathbb{R}^{c \times WH}$, respectively. Given the estimated HR-HS $\mathbf{\tilde{X}}$, the loss function can be defined as:
\begin{equation}\label{eq:loss_hs}
	\mathcal{L}\left(\mathbf{X}, \mathbf{Y}, \mathbf{G}\right) = \mu\Vert\mathbf{\tilde{X}}\mathbf{S} - \mathbf{\tilde{Y}}\Vert_F^2 + \Vert\mathbf{R}\mathbf{\tilde{X}} - \mathbf{\tilde{G}}\Vert_F^2,
\end{equation}
where $\Vert \cdot \Vert_F$ is the Frobenius norm, $\mathbf{S}$ is the spatial downsampling with blurring and $\mathbf{R}$ is the spectral response function that integrates the spectra into R, G, B channels. The first term encourages the spectral similarity between the spatially downsampled $\mathbf{X}$ and $\mathbf{Y}$. The second term encourages the spatial similarity between the spectrally downsampled $\mathbf{X}$ and $\mathbf{G}$. $\mu$ is a scalar controlling the balance between the two terms. The loss function has been widely used with the handcrafted priors in the HS super-resolution~\cite{Lanaras2015,yokoya2012coupled} because the optimization problem is highly ill-posed. Our approach differs from those used in previous studies because it uses GDD as a regularizer.
\subsubsection{Pansharpening.}
Like HS super-resolution, pansharpening also considers two terms that balance the tradeoff between spatial and spectral information. Although the first term in~(\ref{eq:loss_hs}) can be also used for the loss function of pansharpening, the second term may not be effective. This is because the spectral response function of the pansharpening image may partially cover the spectral range captured by the MS image. Thus, the second term cannot effectively measure the spatial similarity between panchromatic and MS images. To address the problem, the second term measuring the spatial similarity is defined as follows:
\begin{equation}
	\mathcal{L}\left(\mathbf{X}, \mathbf{Y}, \mathbf{G}\right) = \mu\Vert\mathbf{\tilde{X}}\mathbf{S} - \mathbf{\tilde{Y}}\Vert_F^2 + \vert\mathbf{D}\triangledown\mathbf{\tilde{X}} - \triangledown\mathbf{\tilde{G}}\vert,
\end{equation}
where $\mathbf{\tilde{Y}}$ is the MS image, $\mathbf{\tilde{G}}$ is the panchromatic image expanded to the same number of bands of $\mathbf{\tilde{X}}$, $\triangledown\mathbf{\tilde{X}}$ is the image gradient of $\mathbf{\tilde{X}}$, $\triangledown\mathbf{\tilde{G}}$ is the image gradient of $\mathbf{\tilde{G}}$, $\vert \cdot \vert$ is the $l_1$ norm, and $\mathbf{D}$ is the diagonal matrix to weight each channel of $\triangledown\mathbf{\tilde{X}}$ so that the magnitude of $\mathbf{\tilde{X}}$ is scaled to that of $\triangledown\mathbf{\tilde{G}}$. Note that $\mathbf{D}$ can be learned with other parameters within the GDD optimization framework. The $l_1$ norm is chosen because this norm more explicitly encourages the edges of the output and guidance images to be similar than other norms (e.g., $l_2$ norm). The first term encourages the spectral similarity while the second term promotes the spatial similarity. A similar loss function has been also explored in~\cite{chen2015sirf}. 
\subsubsection{Denoising.}
For the denoising of the no-flash image, the following loss function was used:
\begin{equation}
	\mathcal{L}\left(\mathbf{X}, \mathbf{Y}\right) = \Vert\mathbf{\tilde{X}} - \mathbf{\tilde{Y}}\Vert_F^2,
\end{equation}
where $\mathbf{\tilde{Y}}$ is the no-flash image. Only $\mathbf{\tilde{X}}, \mathbf{\tilde{Y}}$ are considered in the loss function. $\mathbf{\tilde{G}}$ is considered only in the network architecture because in the detail transfer of the flash and no-flash images, the spatial structures or colors are not necessarily consistent~\cite{Petschnigg2004}. To fairly compare the results derived by DIP~\cite{Ulyan2018}, we adopt the same loss function.

Different handcrafted priors are usually considered with task-specific loss functions. As a result, an optimization framework can be also different for each task. Our approach is different from the previous studies in that GDD is used as a common prior for all of the tasks in a unified optimization framework.
\begin{figure}[t!]
        \begin{subfigure}[b]{0.28\columnwidth}
                \centering
                \includegraphics[width=\linewidth]{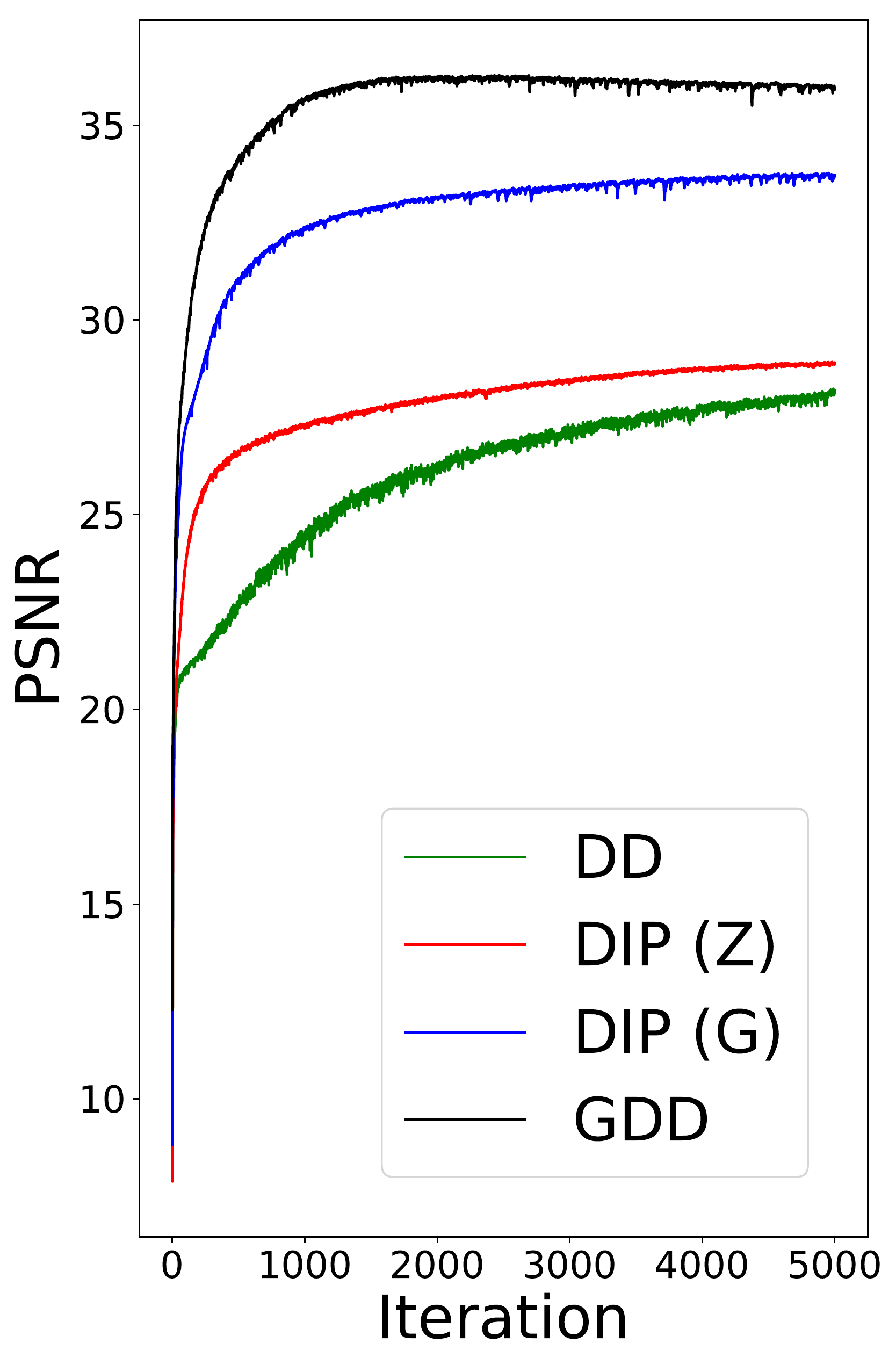}
        \end{subfigure}%
        \begin{subfigure}[b]{0.72\columnwidth}
                \centering
                \includegraphics[width=\linewidth]{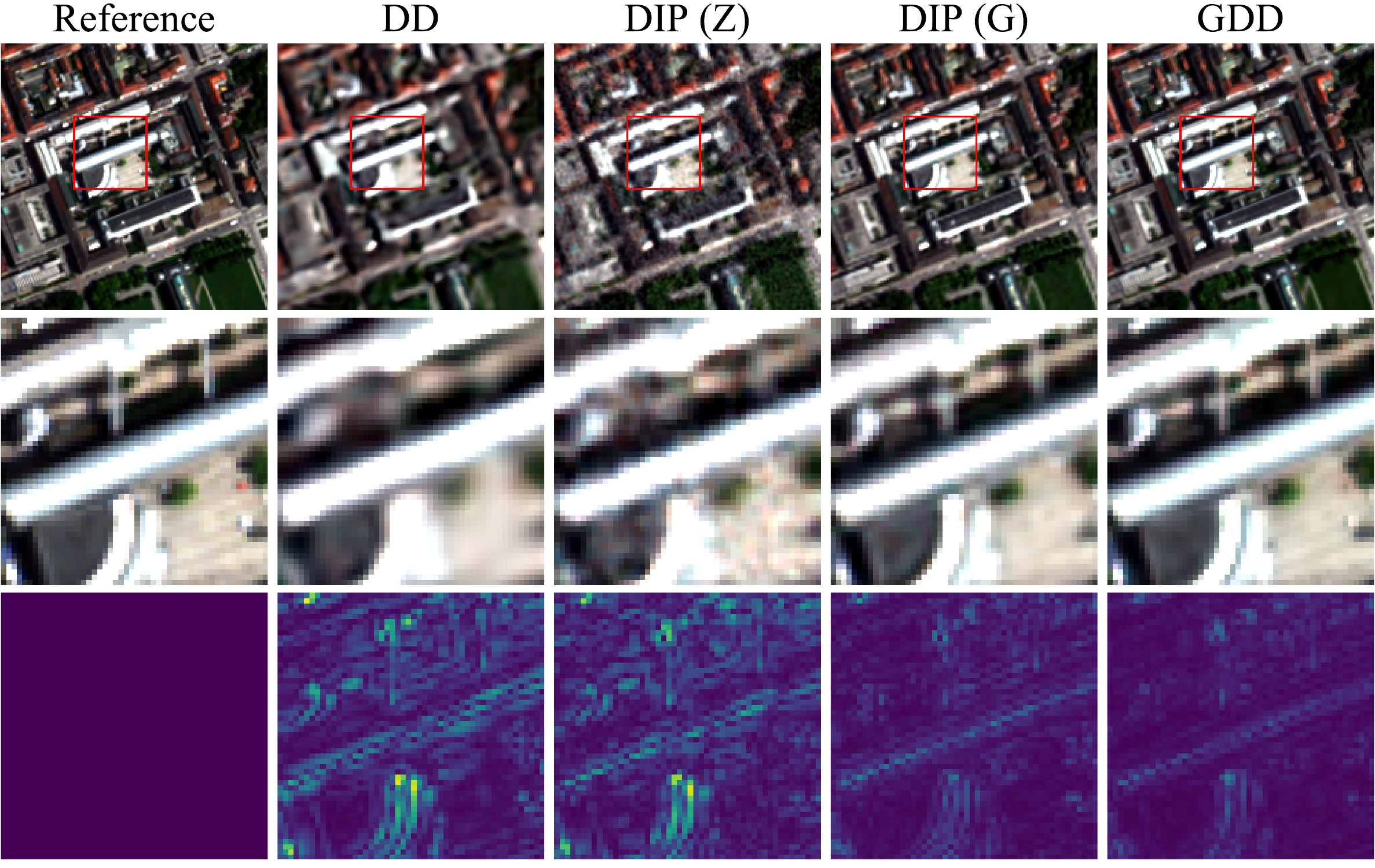}
        \end{subfigure}
        \caption{Comparison of DD, DIP, and GDD. The left figure shows PSNR at different iterations. The right figure shows the images derived at the 5000 iterations. From top to bottom, RGB images, enlarged RGB images, and the error maps of the compared methods.}\label{fig:comparison}
\end{figure}
\section{Comparison between DD, DIP, and GDD}
In this section, we show the comparison between a deep decoder (DD), DIP, and GDD to discuss how GDD outperforms the compared methods. The pansharpening problem was chosen to evaluate the methods as an example. Extensive experiments, including other applications, are shown in the following section. Fig.~\ref{fig:comparison} shows peak signal-to-noise ratio (PSNR) at different iterations. DD uses a tensor representing random noise as an input. DD corresponds to the deep decoder part in GDD. DD is considered for comparison to validate whether the features guided by the encoder-decoder network are really useful. DIP~(Z) represents the deep image prior that uses a random tensor as an input, while DIP~(G) uses a guidance image (i.e., panchromatic imagery) as an input in the encoder-decoder network. \rev{Because DD considers only the decoder part, the information lost in the process of upsampling cannot be recovered. DIP(Z) can use the features derived by a skip connection as a bias term and try to compensate for the lost information. This led to slightly better results of DIP(Z).} GDD and DIP~(G) that incorporate the guidance image produced high PSNR at early iterations. This shows that the use of the guidance image leads to the high quality of the HR-MS image at fewer iterations. Although both GDD and DIP~(G) use the guidance image, GDD considerably outperformed DIP in terms of PSNR. Fig.~\ref{fig:comparison} also shows the RGB images of the reconstructed images, the enlarged RGB, and the corresponding error maps. The enlarged RGB image derived from DD is blurred. The image derived by DIP~(Z) is also blurred and the texture is not correctly recovered. In the highly ill-posed optimization problem, the deep prior that does not incorporate the guidance image cannot produce satisfactory results. DIP~(G) performs better than DD or DIP~(Z). However, the small objects or boundaries of the image are missing in the reconstructed image. GDD preserved the small objects or boundaries more explicitly than DIP~(G), which led to smaller errors. In addition, GDD produced smaller errors in the homogeneous regions of the objects.
\subsubsection{Reasons why GDD is a good regularizer.}
\begin{figure*}[!t]
        \centering
        \includegraphics[width=\linewidth]{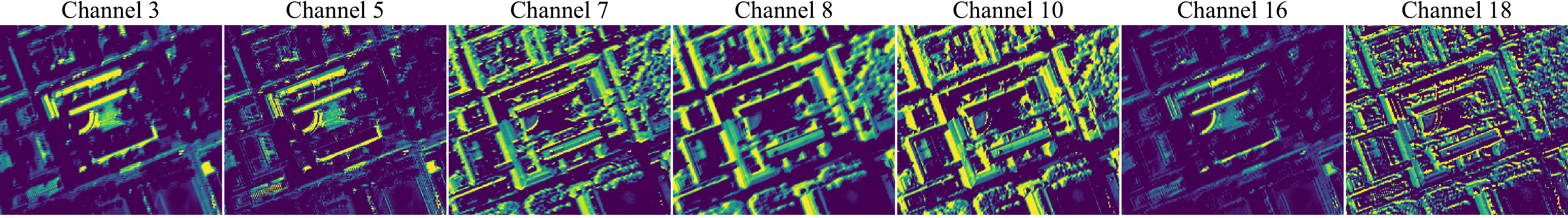}
        \caption{Examples of the conditional weights of different channels used in the attention gates.}\label{fig:attention}
\end{figure*}

We argue that GDD works as a better regularizer than DIP~(G) for the following two reasons:\\
1. \textbf{Upsampling refinement}: The bilinear upsampling used in DIP and GDD causes a strong bias to promote piecewise smoothness and tends to wash away the small objects or boundaries. GDD differs from DIP because it uses an attention gate to weight the features derived by the upsampling. The attention gate enables the small objects or boundaries to be aware by the conditional weights shown in Fig.~\ref{fig:attention}. Owing to attention gates, GDD can reconstruct spatial details.\\
2. \textbf{Feature refinement at multiple scales}: DIP uses the guidance image as an input in the hourglass architecture. In DIP, the features of each layer in the decoder part of the architecture are conditioned using only the features of the previous layer. GDD enables the features of each layer in the decoder part to be conditioned on the semantic features from the guidance image at multiple scales. The attention gates at multiple scales emphasize salient features within each layer, leading to the semantic alignment between the output image and the guidance image.
\section{Experiments}
In this section, we show how GDD works as a regularizer for different image fusion problems. Because of the limited space, only the selected results are shown in the main document. Additional results are shown in the supplementary material. The network architecture of GDD has been fixed for all of the following experiments to validate the robustness of GDD as a regularizer. It is possible to carefully tune the network architecture for each task. However, we believe that the fixed network architecture that works well for different tasks is more important than a carefully tuned architecture that obtains the best performance only for a specific task. In the following experiments, DIP used the guidance image as an input and the same loss function with GDD for fair comparison. 
\subsection{Hyperspectral super-resolution}
\subsubsection{Dataset.}
The CAVE dataset\footnote{\url{http://www1.cs.columbia.edu/CAVE/databases/}} was chosen for the experiments because it has been extensively used to evaluate HS super-resolution methods~\cite{Dian2018b,qu2018unsupervised,Fu2019hyper,xie2019multispectral}. The CAVE dataset consists of HR-HS images that were acquired in 32 indoor scenes with controlled illumination. Each HR-HS image has the spatial size of $256 \times 256$ with 31 bands representing the reflectance of materials in the scene. We followed the experimental setup of~\cite{xie2019multispectral}, $i.e.,$ the generation of the LR-HS image from the HR-HS image by averaging over $32 \times 32$ pixel blocks and the generation of the RGB image by spectral downsampling on the basis of the spectral response function. The proposed GDD does not require training data. However, for fair comparison with the supervised DL method, we chose 12 images for the test, and the rest of the images were used for training as done in~\cite{xie2019multispectral}.
\begin{figure*}[t!]
        \centering
        \includegraphics[width=\linewidth]{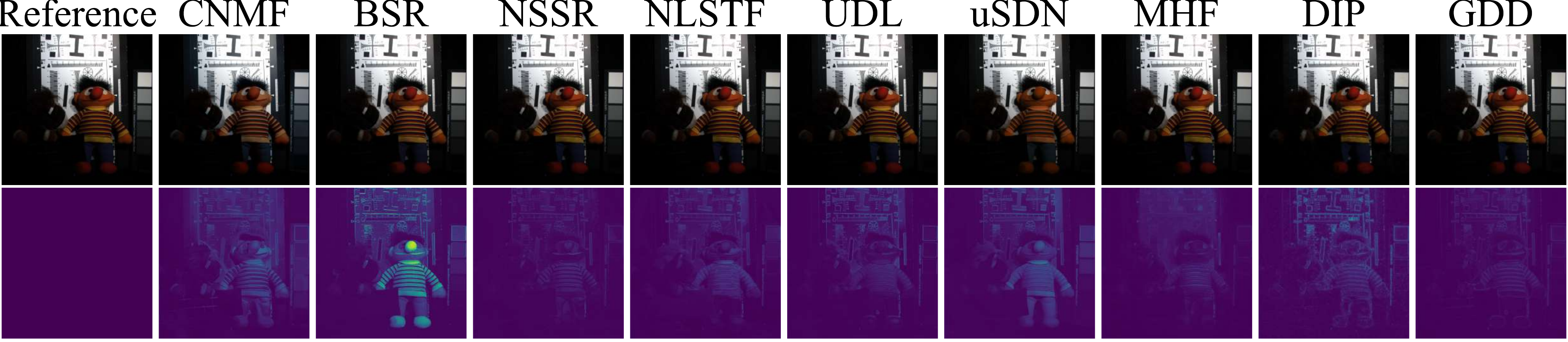}
        \caption{First row: reference and RGB images of the reconstructed HS. The selected results are from \textit{chart and staffed toy} in the CAVE data. Second row: The corresponding error maps.}\label{fig:hsi_result}
\end{figure*}
\subsubsection{Compared methods.}
The compared SOTA methods include the matrix/tensor related methods (CNMF~\cite{yokoya2012coupled}, BSR~\cite{akhtar2015cvpr}, NSSR~\cite{Dong2016} and NLSTF~\cite{Dian2018b}), the supervised DL method (MHF~\cite{xie2019multispectral}), and the unsupervised DL methods (UDL~\cite{Fu2019hyper}, uSDN~\cite{qu2018unsupervised} and DIP~\cite{Ulyan2018}). Among all methods, only MHF required training data.

To quantitatively validate the results, four different criteria were used. The criteria are the root mean square error (RMSE), spectral angle (SA), structural similarity (SSIM~\cite{wang2004image}), and the relative dimensionless global error in synthesis (ERGAS~\cite{wald2000quality}).
\subsubsection{Results.}
 Table~\ref{table:hsi_results} shows the average results of all test images. The performance of BSR and uSDN was worse than those of other methods because the two methods do not assume that the downsampling matrix is available \textit{a priori}. GDD outperformed other unsupervised HS super-resolution methods and was even competitive with the trained DL method (i.e., MHF). This shows that the proposed network architecture is an effective regularizer for the HS super-resolution problem.  Fig. \ref{fig:hsi_result} shows the RGB images of the reconstructed HS images and the error maps. In general, GDD produced lower errors than other methods. The noticeable difference between DIP and GDD is that the errors of DIP are significantly larger at the edges of the image than those of GDD. This implies that GDD properly incorporates the spatial details or semantic features of the guidance image, leading to the edge-preserving image.
 \begin{table}
\centering
\caption{Quantitative results of different metrics on the CAVE dataset. \rev{$\downarrow$ shows lower is better while $\uparrow$ shows higher is better.}}
\label{table:hsi_results}
\begin{tabular}{c|ccccccccc}
\toprule									
		&CNMF	&BSR	&NLSTF &NSSR	&UDL &uSDN	&MHF	&DIP	&GDD	\\
\midrule									
	RMSE$\downarrow$	&3.4557	&5.2030	&2.9414 &2.4247	&2.7971 &4.9289	&2.0827	&3.1589	&\textbf{2.0213}	\\
	ERGAS$\downarrow$	&0.5347	&0.7318	&0.4144 &0.3696	&0.3650 &0.7723	&0.3062	&0.4597	&\textbf{0.3041}	\\
	SA$\downarrow$	    &7.0801	&13.1719 &8.9825 &7.4138 &6.9816 &12.4995 &6.0100 &7.6734 &\textbf{5.5740}	\\
	SSIM$\uparrow$	&0.9760	&0.9524	&0.9805 &0.9770 &0.9733 &0.9385	&\textbf{0.9874} &0.9621 &0.9869	\\
\bottomrule									
\end{tabular}
\end{table}
\begin{figure*}[t!]
        \centering
        \includegraphics[width=\linewidth]{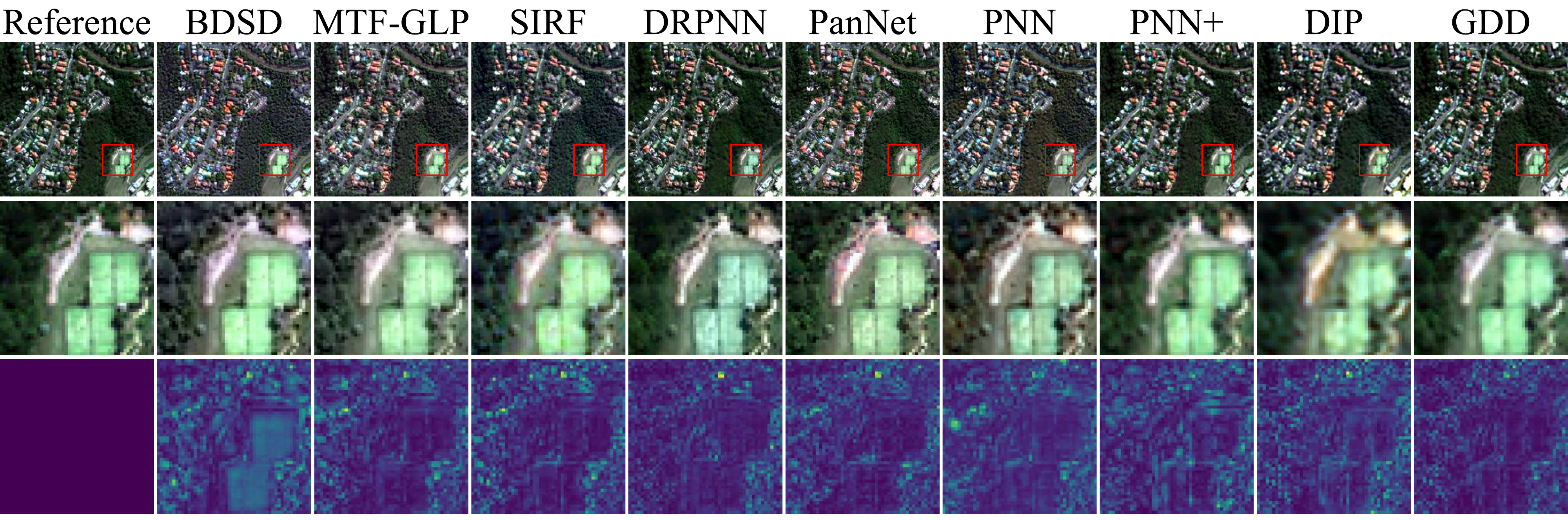}
        \caption{First row: RGB images of the pansharpened MS images. Second row: The enlarged RGB images. Third row: The corresponding error maps.}\label{fig:pan_result}
\end{figure*}
\subsection{Pansharpening}
\subsubsection{Dataset.}
Four different image scenes covering agriculture, urban, forest or mixtures of these were chosen for the experiments. The images were acquired by the WorldView-2. Each MS image is composed of 8 bands representing spectral reflectance. The spatial resolution of the MS image is 2~$m$ while that of the panchromatic image is 0.5~$m$. Each panchromatic image has one band that partially covers the spectral range of the MS image. Synthetic MS and panchromatic images were generated by spatially downsampling the original resolution MS and panchromatic images by the factor of 4. Bicubic downsampling was used. The original resolution MS image was used as reference data. This is the common approach called Wald's protocol~\cite{Wald1997fusion} to generate reference data because reference data (i.e., HR-MS image) are not available~\cite{Fu2019pan,Yang2017}.
\subsubsection{Compared methods.}
The compared SOTA methods include three unsupervised pansharpening methods (BDSD~\cite{garzelli2007optimal}, MTF-GLP~\cite{vivone2014critical}, SIRF~\cite{chen2015sirf}) and four supervised DL methods (DRPNN~\cite{Wei2017}, PanNet~\cite{Yang2017}, PNN~\cite{Masi2016}, PNN+~\cite{Scarpa2018}). The supervised DL methods achieved the SOTA performance. However, the generalization performance of the supervised DL methods is still limited if training data are acquired by a different sensor or in different regions. Training data must be carefully prepared for the supervised DL methods. In this study, we divided each image scene into training and test data acquired by the same sensor. This produces a favorable condition for the supervised DL methods and can be used to validate whether the unsupervised GDD can be comparable to the supervised DL methods.

To qualitatively validate the performance of the methods, the synthetic data (i.e., reduced spatial resolution images) and real data (i.e., original spatial resolution images) were used. Four different criteria were used for evaluation. Similar to the experiments of the HS super-resolution, ERDAS and SA were also considered in pansharpening. In addition, the eight-band extension of average universal image quality index (Q8~\cite{wang2002universal,alparone2004global}), and spatial correlation coefficient (SCC~\cite{zhou1998wavelet}) were used for evaluation. In pansharpening, there are also criteria to validate the performance of the methods on the original spatial resolution images without using reference data. The criteria include a spectral quality index ($\mathbf{D}_{\lambda}$) and a spatial quality index ($\mathbf{D}_{S}$), and the joint spectral and spatial quality with no reference (QNR~\cite{alparone2008multispectral}). The criteria were used to validate the methods using real data (i.e., original spatial resolution of images).
\subsubsection{Results.}
Table~\ref{table:pan_results} shows the average results of all test images. When using the synthetic data with reference data, GDD outperformed other existing methods in terms of all criteria. This showed that GDD reconstructed an HR-MS image that has better quality of both spectral and spatial information. Fig. \ref{fig:pan_result} shows RGB of the reconstructed MS images, the enlarged RGB images, and the corresponding error maps. Although PanNet, PNN, or DRPNN generated sharp edges in the reconstructed images, the spectral information was distorted, which led to the colors that are different from the reference. DIP produced blurred results especially at the edges of the reconstructed images. GDD preserved the spectral information while producing similar spatial details with reference data. This led to lower errors in the reconstructed image. Real images (original resolution images) were also used to evaluate the reconstructed images, as shown in Table~\ref{table:pan_results}. DIP produced the lowest value in terms of D$_\lambda$. This shows that the spectra reconstructed by DIP are most similar to the spectra of the LR-MS image. PNN+ produced the lowest value in terms of D$_s$. This shows that the spatial details reconstructed by PNN+ are the most similar to the spatial details of the pansharpening image. GDD performed better than the other methods in terms of QNR. GDD properly balanced the tradeoff between spectral and spatial resolution, which led to the better value of QNR.
\begin{table}
\vspace{-5mm}
\centering
\caption{Average results of different image scenes for pansharpening. Synthetic represents evaluation with reference at lower resolution. Real represents evaluation with no reference at original resolution. \rev{$\downarrow$ shows lower is better while $\uparrow$ shows higher is better.}}
\label{table:pan_results}
\resizebox{\columnwidth}{!}{
\begin{tabular}{c|c|ccccccccc}											
\toprule											
	&	&BDSD	&MTF-GLP	&SIRF	&DRPNN	&PanNet	&PNN	&PNN+	&DIP	&GDD	\\
\midrule											
\multirow{3}{*}{Synthetic}	&Q8$\uparrow$	&0.8879	&0.9074	&0.8935	&0.9144	&0.9164	&0.9073	&0.9231	&0.9171	&\textbf{0.9469}	\\
	&SA$\downarrow$	&5.9425	&5.4838	&5.9248	&5.3690	&5.4475	&6.5587	&5.7963	&4.6514	&\textbf{4.0254}	\\
	&ERGAS$\downarrow$	&4.6554	&4.1339	&3.9836	&3.6549	&3.9762	&4.1547	&3.7432	&3.5274	&\textbf{2.6879}	\\
	&SCC$\uparrow$	&0.9071	&0.9021	&0.8970	&0.9316	&0.8868	&0.9131	&0.9048	&0.8965	&\textbf{0.9418}	\\
\midrule											
\multirow{3}{*}{Real}	&QNR$\uparrow$	&0.9077	&0.9157	&0.9071	&0.8648	&0.8833	&0.9253	&0.9492	&0.9446	&\textbf{0.9517}	\\
	&D$_\lambda$$\downarrow$	&0.0423	&0.0391	&0.0538	&0.0320	&0.0574	&0.0316	&0.0250	&\textbf{0.0188}	&0.0202	\\
	&D$_s$$\downarrow$	&0.0531	&0.0469	&0.0414 &0.1066	&0.0629	&0.0447	&\textbf{0.0264}	&0.0374	&0.0288	\\
\bottomrule											
\end{tabular}
}
\end{table}
\begin{figure*}[t!]
        \centering
        \includegraphics[width=\linewidth]{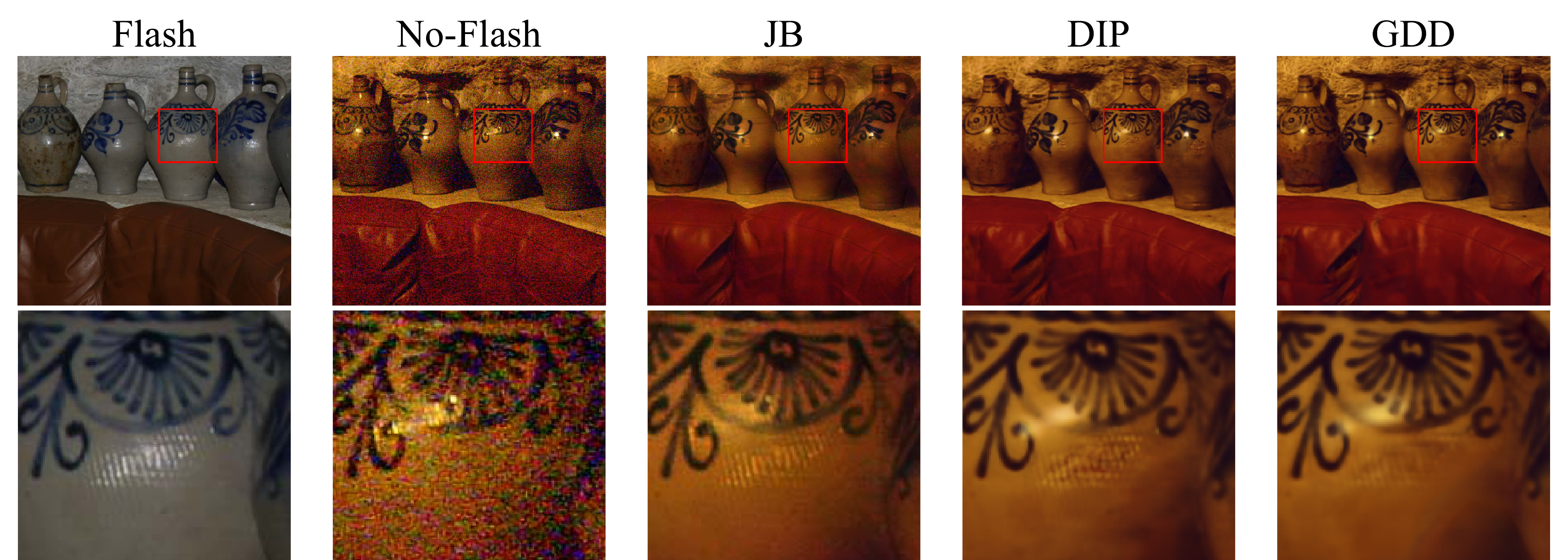}
        \caption{The reconstructed images of the no-flash image with the help of the flash image.}\label{fig:flash_result}
\end{figure*}
\subsection{Denoising}
In this section, the reconstruction of a flash image with the help of a no-flash image was addressed to show another application of GDD. The no-flash image acquires an image under ambient illumination where the image can be noisy because of the low-light conditions~\cite{Petschnigg2004}. However, the flash image acquires an image under artificial light where the image is noise-free and the spatial details of the image are recorded. However, the lighting characteristics are unnatural, and unwanted shadows or artifacts may be produced in the flash image. The objective of this application is to reconstruct a clean no-flash image using the features of a flash image. In this application, true reference data cannot be available. Although an image with long exposure may be used as a reference~\cite{Petschnigg2004}, the magnitude or characteristics of illumination are not necessarily the same as those of the true reference. In this study, the reconstructed images are qualitatively evaluated according to~\cite{Ulyan2018}. In~\cite{Ulyan2018}, DIP that uses the flash image as an input and the no-flash image as an output was successfully applied to the problem. We qualitatively examined if the architecture used in GDD was as effective as DIP.\

Fig.~\ref{fig:flash_result} shows that the reconstructed images of the no-flash image. DIP and GDD removed the artifacts more clearly than the joint bilateral method (JB)~\cite{Petschnigg2004}. GDD produced more explicit boundaries of the image than DIP while preserving the natural colors of the image. This shows that GDD performed at least as well as DIP for the no-flash image reconstruction. 

\section{Conclusion}
We proposed an unsupervised image fusion method that was based on GDD. GDD is a network architecture-based regularizer and can be used to solve different image fusion problems that have been independently studied so far. The network architecture can better exploit spatial details and semantic features of a guidance image. This is achieved by considering an encoder-decoder network that extracts spatial details and semantic features of a guidance image. The multi-scale attention gates enable the extracted semantic features to guide a deep decoder network that generates an output image. This approach achieved the SOTA performance in the different image fusion problems. It pushes the boundaries of the current studies that address only one specific problem. The promising results open up the possibility of a network architecture-based prior that can be used for general purpose including various image fusion problems.
%
%
\bibliographystyle{splncs04}
\bibliography{ref_all}
\end{document}